\documentclass[%
prl,
superscriptaddress,
twocolumn,
nofootinbib,
amsmath,amssymb,aps,
]{revtex4-2}

\usepackage{graphicx}% Include figure files
\usepackage{dcolumn}% Align table columns on decimal point
\usepackage{bm}% bold math
\usepackage{comment}
\usepackage{tikz}
\usepackage{pgfplots}
\pgfplotsset{compat=newest}
\usepackage{amsmath}
\usepackage{amsthm}
\usepackage{physics}
\usepackage{mathtools}
\usepackage{amsmath}
\usepackage{amsthm}
\usepackage{physics}
\usepackage[compat=1.1.0]{tikz-feynman}
\usepackage[T1]{fontenc} % if needed
\usepackage{tensor}
\usepackage{graphicx}
\usepackage[export]{adjustbox}
\usepackage{slashed}
\usepackage{stackrel,amssymb}
\usepackage{tikz}
\usepackage{mathtools}
\usetikzlibrary{arrows.meta}
\usepackage{comment}
\tikzset{>={Latex[scale=1.1]}}
\usetikzlibrary{arrows.meta,
	bending,
	decorations.markings, decorations.text}
\newcommand{\ee}{\mathrm{e}}

\newcommand{\vvev}[1]{{\left\langle #1 \right\rangle}}
\makeatletter
\newcommand*{\letterdef@}{}
\newcommand*{\letterdef}[3]{%
	\def\letterdef@##1{\expandafter\newcommand\csname #1\endcsname{#2{##1}}}%
	\@tfor\@tempa :=#3\do{\expandafter\letterdef@\expandafter{\@tempa}}}
\makeatother
\letterdef{c#1} {\mathcal}{ABCDEFGHIJKLMNOPQRSTUVWXYZ} % \cX = \mathcal{X}
\letterdef{rm#1}{\mathrm} {dDeimM} % \rmX = \mathrm{X} for X in {dDmM}

\begin{document}

%\preprint{APS/123-QED}

\title{Supersymmetric localization and non-conformal $\mathcal{N}=2$ SYM theories in the perturbative regime}

\author{Marco Bill\`o}
\email{marco.billo@unito.it}
\affiliation{Universit\`a di Torino, Dipartimento di Fisica and INFN, Sezione di Torino,\\
	Via P. Giuria 1, I-10125 Torino, Italy}
\author{Luca Griguolo}
 \email{luca.griguolo@unipr.it}
 \affiliation{Dipartimento SMFI, Universit\`a di Parma and INFN Gruppo Collegato di Parma,\\ Viale G.P. Usberti 7/A, 43100 Parma, Italy
}
\author{Alessandro Testa}%
 \email{alessandro.testa@unipr.it}
\affiliation{Dipartimento SMFI, Universit\`a di Parma and INFN Gruppo Collegato di Parma,\\ Viale G.P. Usberti 7/A, 43100 Parma, Italy
}

\begin{abstract}
We examine the relation between supersymmetric localization on $\mathbb{S}^4$ and standard QFT results for non-conformal theories
in flat space. Specifically, we consider 1/2 BPS circular Wilson loops in four-dimensional  SU($N$) $\mathcal{N}$= 2 SYM theories with massless hypermultiplets in an arbitrary representation $\mathcal{R}$ such that the $\beta$-function is non-vanishing. On  $\mathbb{S}^4$, localization maps this observable into an interacting matrix model.
Despite broken conformal symmetry at the quantum level, we show that within a specific  regime of validity the matrix model predictions are consistent with perturbation theory in flat space up to order $g^6$. At this order,  the reorganization of  Feynman diagrams based on the matrix model potential, which has been widely tested in conformal models, also applies in non-conformal set-ups and is realized, in perturbative field theory,  through highly non-trivial interference mechanisms.
\end{abstract}
\pacs{...}
\maketitle

%\tableofcontents

\section{Introduction}
Localization techniques have represented a major breakthrough 
in the study of supersymmetric gauge theories on compact manifolds at non-perturbative level \cite{Pestun:2007rz}. Several exact results have been obtained for partition functions \cite{Kapustin:2009kz,Jafferis:2010un,Benini:2012ui,Doroud:2012xw,Kim:2012ava}, Wilson loops \cite{Erickson:2000af,Drukker:2000rr,Drukker:2007qr,Drukker:2010nc,Drukker:2009hy,Pestun:2009nn,Bianchi:2018bke} , line defects \cite{Pestun:2009nn,Gomis:2011pf,Kapustin:2012iw,Drukker:2012sr} and other supersymmetric observables \cite{Giombi:2009ds,Gerchkovitz:2016gxx,Giombi:2018qox,Komatsu:2020sup,Chester:2019jas,Dorigoni:2021guq}, enabling non-trivial checks of the AdS/CFT duality \cite{Buchel:2013id,Benini:2015eyy,Cabo-Bizet:2018ehj,Binder:2019jwn} also in non-maximally supersymmetric models. 
	
Technically, the finite volume of spacetime plays an essential role in the localization procedure and serves as a (natural) gauge-invariant regulator for IR divergences. Conversely,  the UV structure of the theory is left unchanged by the compactification and generally requires a renormalization.  In superconformal models, the localization predictions naturally extend to the infinite flat space, and it is possible to compare them with standard field theory approaches. This program has been actively conducted in four dimensions, where the matrix model generated by supersymmetric localization on $\mathbb{S}^4$ was successfully tested at weak coupling against perturbative approaches for BPS Wilson loops \cite{Bassetto:2008yf,Andree:2010na,Billo:2019fbi} and special local correlators \cite{Rodriguez-Gomez:2016ijh,Billo:2017glv,Billo:2018oog}. These analyses reveal that perturbative computations in flat space are captured by a one-loop effective action on $\mathbb{S}^4$ \cite{Pestun:2007rz}, which provides an elegant reorganization of the different
Feynman diagrams \cite{Billo:2019fbi}.
		
However, when the gauge theory contains dimensionful parameters, such as a mass term in $\mathcal{N}=2^*$ theories or a scale generated by dimensional transmutation, conformal symmetry in flat space is broken. As a result,  the short and long distance properties of the model are different and the calculation in $\mathbb{R}^4$ and on $\mathbb{S}^4$ are no longer expected to match. In particular, when the theory contains a mass scale,  observables calculated on $\mathbb{S}^4$ acquire a dependence on the dimensionless parameter constructed by the product of the mass scale and of the radius of the sphere, leading to different results with respect to the flat space.   This scenario was examined in \cite{Belitsky:2020hzs} studying the vacuum expectation value of the half-BPS circular Wilson loop in $\mathcal{N}=2^*$ SYM and finding that the perturbative two-loop computation of the observable on $\mathbb{S}^4$ coincides with the localization result, while the analogous flat-space computation differs. 
		 
While a mass deformation violates the conformal invariance at classical level and affects both the structure of the propagators and of the action \cite{Bobev:2013cja,Festuccia:2011ws}, the presence of a non-vanishing beta-function 
%in theories with a classical conformal symmetry 
yields a breaking of the map between $\mathbb{S}^4$ and $\mathbb{R}^4$ at the \textit{quantum} level. However, supersymmetric localization still provides explicit expressions in terms of a one-loop exact running coupling constant \cite{Pestun:2007rz}, in analogy to the flat-space computations. Therefore, it is natural to investigate if the conventional perturbative series, when expressed in terms of the running coupling, is encoded in the localization effective action or to understand which part of that (if any) is univocally contained. 

In this letter, we consider SU($N$) $\mathcal{N}=2$ SYM theories defined in flat space with massless hypermultiplets in a generic representation $\cR$. The $\mathcal{N}=2$ vector multiplet contains the gauge field $A_\mu$, a complex scalar $\phi$  and two Weyl fermions, while the hypermultiplets are described by two complex scalars along with their fermionic superpartners. In these theories, the $\beta$-function is one-loop exact, i.e.
\begin{equation}
	\label{beta0}
	\beta(g)=\beta_0 g^3 \ , \quad \text{where} 	\quad	\beta_0 =\frac{i_\mathcal{R}-N}{8\pi^2}~ .
\end{equation}			
In the previous expression, we denoted with   $i_\mathcal{R}$ the Dynkin index\footnote{The Dynkin index is defined by $\Tr_\mathcal{R} T^a T^b=i_\mathcal{R} \delta^{ab}$, with the normalization that $i_F=1/2$ for the fundamental representation.} of the representation $\mathcal{R}$. Throughout this work, we will consider  asymptotically free theories, where $\beta_0 <0$. We will focus on the half-BPS circular Wilson loop operator in the fundamental representation
\begin{equation}
	\label{eq:1/2 BPS Wilson loop}
	\widehat{W} = \dfrac{1}{N} \text{tr} \ \mathcal{P} \exp \bigg\{ g_B \int_C \dd{\tau} \bigg[ i
	A_\mu \dot x^\mu + \dfrac{R}{\sqrt{2}} \Big( \phi +\phi^\dagger \Big)\bigg]
%\dd A(\tau) + \dfrac{R}{\sqrt{2}} \Big( \phi +\phi^\dagger \Big)(\tau)\bigg]
	\bigg\} \ ,
\end{equation} where $g_B$ is the bare coupling constant, $\cP$ is the path-ordering operator
%, while the gauge field and vector-multiplet scalar are 
and the integral is over a circle $C$ of radius $R$ parametrized as $x_\mu(\tau)=R(\cos\tau,\sin\tau,\mathbf{0})$. 

In the following, we will show that the matrix model predictions match standard 
perturbation theory up to three-loop accuracy\footnote{The full technical details of the Feynman diagram computations will be given in a upcoming paper \cite{Billo:2024}} \textit{within} a specific range of validity (see eq. (\ref{eq:range})).
Specifically,  supersymmetric localization predicts two corrections proportional to $\zeta(3)$ which, in perturbative field theory, possess a different origin: one of them is present also in superconformal cases \cite{Andree:2010na,Billo:2017glv,Billo:2019fbi} and arises from a Feynman integral which has the same form on $\mathbb{R}^4$ and $\mathbb{S}^4$, 
while the second one emerges by interference effects between evanescent terms and the UV divergence of the bare coupling constant.

\section{BPS Wilson loop on $\mathbb{S}^4$ via localization}
Compactifying the theory on a four-sphere of radius $R$ and for which $C$ is an equator, supersymmetric localization \cite{Pestun:2007rz} enables to compute the expectation value of the half-BPS Wilson loop defined in eq. (\ref{eq:1/2 BPS Wilson loop}) by an integral over a traceless Hermitian $N\times N$ matrix $a$:
\begin{align}
	\label{eq:ZN=2generic}
	\cW = \dfrac{1}{\cZ} \int \dd a\, \mathrm{e}^{- \tr a^2 - S_{\text{int}}(a,g)}\,\widehat{\cW}(a)
	~,
\end{align} 
where the matrix operator $\widehat{\cW}$ reads
\begin{equation}
	\widehat{\cW} =\dfrac{1}{N}\tr\exp(\dfrac{g a}{\sqrt{2}}) = 1 + \frac{g^2}{4N} \tr a^2 + \cO(g^4)
\end{equation}
and the partition function $\cZ$ is given by the same integral without the insertion of $\widehat{\cW}$. 
The integration measure is such that  $\int \dd a  \ \mathrm{e}^{-\tr a ^2}=1$. If we neglect the instanton contributions, the interaction potential  $S_{\mathrm{int}}(a,g)$ arises
from the one-loop determinants around the fixed point of the localizing action and  is given by 
\cite{Billo:2023igr}
\begin{align}
	\label{eq:interaction action}
	S_{\text{int}}(a,g) & =
	- \sum_{m=2}^{\infty}\left(-\frac{g^2}{8\pi^2}\right)^m \dfrac{\zeta(2m-1)}{m} 
	\Tr_\mathcal{R}^\prime a^{2m}~, 
\end{align}
where $\Tr_\mathcal{R}^\prime \bullet=\left(\Tr_{\mathcal{R}}\bullet-\Tr_{\text{Adj}}\bullet\right)$. This combination of traces describes the matter content of the difference between the $\mathcal{N}=2$ models under consideration and $\mathcal{N}=4$ SYM. From the perspective of perturbative field theory, eq. (\ref{eq:interaction action}) suggests constructing the interaction contributions by considering  diagrams with internal lines in representation $\mathcal{R}$, and then subtracting identical contributions in which $\mathcal{R}=\mathrm{Adj}$. Diagrammatically, we will  depict the corrections characterized by matter in the so-called ${\it difference \  theory}$ \cite{Andree:2010na,Billo:2019fbi} by a double solid/dashed line.  For instance, the expected correspondence between a contribution in the matrix model involving the quartic  vertex $\Tr_{\mathcal{R}}^\prime a^4$ and usual Feynman diagrams is 
\begin{equation}
	\label{eq:difference theory}
	\Tr_{\mathcal{R}}^\prime a^4 = \mathord{
		\begin{tikzpicture}[radius=2.cm, baseline=-0.65ex,scale=0.5]
			\begin{feynman}
				\vertex (A) at (1,0);
				\vertex (B) at (-1,0);
				\vertex (C) at (0,1);
				\vertex (D) at (0,-1);
				\vertex (O) at (0,0);
				\diagram*{
					(A) --[ plain,thick] (O),
					(D) --[ plain,thick] (O),
					(B)--[plain,thick] (O),
					(C)--[plain,thick] (O),
				};
			\end{feynman}
			\node at (0,0)[circle,fill,inner sep=2pt]{};
	\end{tikzpicture} } 
 \ \quad  \leftrightarrow \quad  
%	\text{vertex} \quad  \leftrightarrow \quad  
	\mathord{
		\begin{tikzpicture}[radius=2.cm, baseline=-0.65ex,scale=0.55]
			%\filldraw[color=gray!80, fill=gray!15](0,0) circle (1);	
			\draw [black, thick]   (0,0) circle [radius=1.];
			\draw [black, thick, dashed]   (0,0) circle [radius=0.9];
			\begin{feynman}
				\vertex (A) at (0,2.);
				\vertex (C) at (0, 1.);
				\vertex (D) at (0, -1.);
				\vertex (d) at (0.1,0) ;
				\vertex (B) at (0,-2.);
				\vertex (a) at (-1,0);
				\vertex (a1) at (-2,0);
				\vertex (b) at (1,0);
				\vertex (b1) at (2,0);
				\diagram*{
					(A) -- [photon] (C),
					(A) --[plain] (C),
					(D) -- [photon] (B),
					(D) --[plain] (B),
					(a1)--[photon] (a),
					(a1)--[plain] (a),
					(b)--[photon] (b1),
					(b)--[plain] (b1),
				};
			\end{feynman}
	\end{tikzpicture} } \ ,
\end{equation}
where the superposition of a  wiggly/straight line denotes the vector-multiplet field propagation.

Importantly, in (\ref{eq:ZN=2generic}) and (\ref{eq:interaction action}), $g=g(R)$ is the running coupling constant  
\begin{equation}
	\label{eq:running coupling}
	\frac{1}{{g}^2} = \frac{1}{g_*^2} +\beta_0 \log M^2R^2~ , 
\end{equation} 
where $g_*=g_*(M)$ is the renormalized coupling at the UV cut-off $M$. This scale enters the matrix model because the representation $\cR$ is associated with a non-vanishing $\beta$-function. This requires a regularization for the one-loop fluctuation determinants which involves additional hypermultiplets of mass $M$, see \cite{Pestun:2007rz,Billo:2023igr}. 
For perturbation theory to applicable,  asymptotic freedom requires that
\begin{equation}
	\label{eq:range}
	\Lambda \ll\frac{1}{R}\ll M  \ , \quad \text{where} \quad	\Lambda = M \ee^{\frac{1}{2g_*^2\beta_0}}~.
\end{equation} 
 is the \textit{infrared} strong coupling scale generated by dimensional transmutation.
%\begin{equation}
%	\Lambda = M \ee^{\frac{1}{2g_*^2\beta_0}}~.
%\end{equation} 
Indeed, when $1/R$ approaches $\Lambda$, the running coupling $g$ is of order $\cO(1)$, requiring a resummation of the perturbative series, and the observable also receives non-perturbative power-like corrections\footnote{
In certain multicolour models, such as $\mathcal{N}=2^*$ or the massive deformation of $\mathcal{N}=2$ SQCD, the coefficients $C_n$ were determined on $\mathbb{S}^4$ by localization techniques \cite{Russo:2013kea}. Instantons, which we neglected in the matrix model, would also contribute to the observable with terms of this type.} $C_n(R\Lambda)^n$. 
We expect  that such \textit{infrared} contributions differ between the sphere and flat space due to the  conformal anomaly.	Conversely, when $1/R$ approaches $M$,  the 
(massive) regulating degrees of freedom become relevant and the theory itself changes.

The matrix model (\ref{eq:ZN=2generic}) is \textit{formally} analogous to that employed in the conformal case in \cite{Billo:2019fbi}, so that we can apply the same techniques for the perturbative calculation.  Up to order  $g^6$, the prediction is 
\begin{equation}
	\label{eq:localization prediction}
\begin{split}
	\cW &= \cW_0 
		+ \frac{g^6\,\zeta(3)}{2^9\pi^4 N} \vvev{\tr a^2\,\Tr_\mathcal{R}^\prime a^4}_{0,c}	
+ \cO(g^8) \\[0.3em]
 &= \cW_0 + \dfrac{3g^6\zeta(3)}{2^8\pi^4N}\cK_4^{\prime} + \dfrac{g^6\zeta(3)}{16\pi^2}C_FN\beta_0 +\cO(g^8) \ ,
\end{split}
\end{equation} 
where the subscript $0,c$  denotes the connected correlator in the Gaussian matrix model, while $C_F=(N^2-1)/2N$ is the fundamental Casimir. Moreover, $\cW_0$ captures the Wilson loop expectation value in the free matrix model  and, in $\mathcal{N}=4$ SYM, it resums all the \textit{ladder-like} diagrams. Its explicit expression reads
\cite{Drukker:2000rr,Erickson:2000af}
\begin{align}
	\label{W0is}
		\cW_0 & = \dfrac{1}{N} L^1_{N-1}(-g^2/4)\,\ee^{\frac{{g}^2}{8}(1-1/N)}
		= 1 + \frac{g^2}{4} C_F \notag\\
		& +  \frac{g^4C_F (2N^2-3)}{192 N}
		+  \frac{g^6 C_F(N^4-3N^2+3) }{4608 N^2}+ \cO(g^8)~,
\end{align}
where $L_{N-1}^1$ is a Laguerre polynomial. 

To evaluate the connected correlator for arbitrary $\cR$, we introduced the free contraction $\big<a^a a^b\big>_0=\delta^{ab}$ and employed the usual Wick theorem.  The two interaction terms, characterized by the colour factors $C_F N\beta_0$ and  $\cK_4^\prime=\Tr_{\mathcal{R}}^\prime T^a T^e T_a T_e=2N C_F\left(C_\cR-\frac{Ni_\cR}{2} -\frac{ N^2}{2}\right)$ are associated with the two contractions of the matrix model quartic vertex:
\begin{equation}
\label{eq:matrix model diagrams}
\mathord{
	\begin{tikzpicture}[baseline=-0.65ex,scale=0.5]
		\draw [black] (0,0) circle [radius=1.5cm];
		\begin{feynman}
			\vertex (A) at (0,1.5);
			\vertex (B) at (0,-1.5);
			\vertex (C) at (0,0);
			\vertex (D) at (-1,0);
			\diagram*{
				(A) -- [plain, thick] (B),
				(C)--[plain,thick,half right] (D),
				(C)--[plain,thick,half left] (D),	
			};
		\end{feynman}
		\node at (0,0)[circle,fill,inner sep=2pt]{};
	\end{tikzpicture} 
} \ , \quad \quad \mathord{
	\begin{tikzpicture}[baseline=-0.65ex,scale=0.5]
		\draw [black] (0,0) circle [radius=1.5cm];
		\begin{feynman}
			\vertex (A) at (0,1.5);
			\vertex (B) at (0,-1.5);
			\vertex (C) at (0,-0.5);
			\vertex (D) at (0,0.5);
			\diagram*{
				(A) -- [plain, thick] (B),
				(D)-- [plain,half right,thick] (C),
				(D)-- [plain,half left,thick] (C),
			};
		\end{feynman}
		\node at (0,0.5)[circle,fill,inner sep=2pt]{};
	\end{tikzpicture}  \ .
}
 \end{equation}

The correspondence between matrix vertices and field theory matter loops (\ref{eq:difference theory}) suggests that these interactions should 
arise from single-exchange field theory diagrams.
This connection has been checked%
\footnote{In \cite{Billo:2019fbi}, the test has been extended to four-loop order in generic superconformal set-ups.}
long ago in \cite{Andree:2010na,Billo:2019fbi} for generic superconformal set-ups, where only the $\cK^\prime_4$ structure is present.  However, in non-conformal models it is not obvious if this  correspondence  persists.  

\section{Field theory approach in flat space}
We regularize Feynman diagrams by \textit{dimensionally reducing} the theory to $d=4-2\epsilon$ dimensions with $\epsilon>0$ \cite{Erickson:2000af}. This scheme preserves the extended supersymmetry of the model but breaks classical conformal symmetry since $g_B$ is dimensionful. Consequently, the v.e.v  of the half-BPS Wilson loop operator (\ref{eq:1/2 BPS Wilson loop})
can only depend on the dimensionless combination $\hat g_B = g_B R^\epsilon$:  
\begin{equation}
	\label{eq:perturbative}
\big<\widehat W\big> \equiv W = 1 + \hat g_B^2 W_2 + \hat g_B^4 W_4 + \hat g_B^6 W_6 + \cO(\hat g_B^8) \ .
\end{equation}
\subsection{One-loop corrections}
The one-loop correction $\hat{g}_B^2 W_2$ arises from the following single-exchange diagram \begin{equation}
	\label{eq:ladder g2}
		W_2 = 
		\mathord{
		\begin{tikzpicture}[baseline=-0.65ex,scale=0.4]
			\draw [black] (0,0) circle [radius=2cm];
			\begin{feynman}
				\vertex (A) at (0,2);
				\vertex (B) at (0,-2);
				\diagram*{
					(A) -- [photon] (B),
					(A) --[ fermion] (B),
				};
			\end{feynman}
		\end{tikzpicture} 
	} =  C_F A_1(\epsilon)\ ,
\end{equation} where we used the graphical notation of eq. (\ref{eq:difference theory}) to denote the gauge-field/adjoint-scalar propagation inside the Wilson loop and we defined the functions 
\begin{equation}
	A_n(\epsilon) = \frac 18 \pi^{n\epsilon} \Gamma^n(1 - \epsilon) \dfrac{\sec(n\pi \epsilon)\Gamma(-n\epsilon) }{\Gamma(-2n\epsilon)\Gamma(1+ n\epsilon)} 
	= \frac 14 + \cO(\epsilon)\ .
\end{equation}
Note that 
%in the limit $\epsilon\to 0$, eq. (\ref{eq:ladder g2}) is regular and that, with the convention of eq. (\ref{eq:perturbative}), 
we do not include the factors $\hat g_B$ when we give the explicit result of a diagram.

\subsection{Two-loop corrections}
The Feynman diagrams which contribute at order $\hat{g}_B^4$ are organized in three different classes \cite{Billo:2023igr}
\begin{equation}
	\label{eq:two-loop diagrams}
%\hat{g}_B^4
W_4=\mathord{
		\begin{tikzpicture}[radius=2.cm, baseline=-0.65ex,scale=0.4]
			\draw [black] (0,0) circle [];
			%\filldraw[color=gray!15, fill=gray!15](0,0) circle (1.1);	
			\draw [black, dashed, thick] (0,0) circle [radius=1.1cm];
			\begin{feynman}
				\vertex (A) at (0,2.);
				\vertex (C) at (0, 1.1);
				\vertex (D) at (0, -1.1);
				\vertex (B) at (0,-2.);
%				\vertex (d) at (0.1,0.) {\text{\footnotesize 1-loop }\normalsize} ;
				%			\vertex (d) at (0.1,-0.3) {\text{\footnotesize $\cR$ }\normalsize} ;
				\diagram*{
					(A) -- [photon] (C),
					(A) --[ fermion] (C),
					(D) -- [photon] (B),
					(D) --[fermion] (B)
				};
			\end{feynman},
	\end{tikzpicture} }  \   +  \ 
		\mathord{
		\begin{tikzpicture}[radius=2.cm, scale=0.4, baseline=-0.65ex]
			\draw [black] (0,0) circle [];
			\begin{feynman}
				\vertex (A) at (0,2.);
				\vertex (C) at (0,0);
				\vertex (D) at (-1.5, -1.3);
				\vertex (B) at (1.5, -1.3);
				\diagram*{
					(A) -- [photon] (C),
					(A) --[ fermion] (C),
					(C) -- [photon] (B),
					(C) --[fermion] (D),
					(C) --[ photon] (D)
				};
			\end{feynman} 
		\end{tikzpicture}  \ + \ 
	}  
	\mathord{
		\begin{tikzpicture}[baseline=-0.65ex,scale=0.4]
			\draw [black] (0,0) circle [radius=2cm];
			\begin{feynman}
				\vertex (A) at (-0.75,1.8);
				\vertex (B) at (-0.75,-1.8);
				\vertex (C) at (0.75,1.8);
				\vertex (D) at (0.75,-1.8);
				\diagram*{
					(A) -- [photon] (B),
					(A) --[ fermion] (B),
					(C) -- [photon] (D),
					(C) --[ fermion] (D)
				};
			\end{feynman}
		\end{tikzpicture} 
	}  \ .
\end{equation}
The internal  bubble in the first diagram denotes the one-loop correction to the adjoint scalar and gauge field propagator. Specifically, the dashed line is associated with the matter fields in the representation $\cR$ which run in the virtual loop. This correction, as well as the diagrams with internal vertices, exhibit a (UV) singular behaviour when $\epsilon\to0$. The singularity in the \textit{Mercedes-like} diagrams arises when two points on the contour collide and is such that \cite{Erickson:2000af, Billo:2023igr}
\begin{equation}
	\label{eq:difference}
	\mathord{
		\begin{tikzpicture}[radius=2.cm, scale=0.48, baseline=-0.65ex]
			\draw [black] (0,0) circle [];
			\begin{feynman}
				\vertex (A) at (0,2.);
				\vertex (C) at (0,0);
				\vertex (D) at (-1.5, -1.3);
				\vertex (B) at (1.5, -1.3);
				\diagram*{
					(A) -- [photon] (C),
					(A) --[ fermion] (C),
					(C) -- [photon] (B),
					(C) --[fermion] (D),
					(C) --[ photon] (D)
				};
			\end{feynman} 
		\end{tikzpicture} 
	} = - \mathord{
	\begin{tikzpicture}[radius=2.cm, baseline=-0.65ex,scale=0.48]
		\draw [black] (0,0) circle [];
		%\filldraw[color=gray!80, fill=gray!15](0,0) circle (1);	
		\draw [black,thick]   (0,0) circle [radius=1.];
		\begin{feynman}
			\vertex (A) at (0,2.);
			\vertex (C) at (0, 1.);
			\vertex (D) at (0, -1.);
			\vertex (B) at (0,-2.);
%			\vertex (d) at (0.1,0.) {\text{\footnotesize 1-loop }\normalsize} ;
			%					\vertex (d) at (0.1,-0.3) {\text{\footnotesize $\rm adj$ }\normalsize} ;
			\diagram*{
				(A) -- [photon] (C),
				(A) --[ charged scalar] (C),
				(D) -- [photon] (B),
				(D) --[ charged scalar] (B)
			};
		\end{feynman}
\end{tikzpicture} } -\epsilon\dfrac{\zeta(3)C_F N}{8\pi^2}+\cO(\epsilon^2) \ ,
\end{equation} where the internal bubble on the right-hand side denotes the one-loop correction to the adjoint scalar and gauge field propagator in $\mathcal{N}=4$ SYM, where the hypermultiplets are in the adjoint representation. The previous expression reveals in $\cN=4$ SYM,  all the interaction diagrams cancel each other out and the observable receives contributions only from the ladder-like corrections.
The evanescent $\zeta(3)$-term results from a triple \textit{path-ordered} integration and possesses the same colour factor, proportional to $C_F$, of the single exchange diagrams (\ref{eq:ladder g2}). Upon renormalization, the UV poles of the bare coupling $g_B$  interfere with the evanescent factor, leading to a finite three-loop contribution. 

Substituting eq. (\ref{eq:difference}) in eq. (\ref{eq:two-loop diagrams}), 
we find that 
\begin{equation}
	W_4 = 
	\mathord{
		\begin{tikzpicture}[radius=2.cm, baseline=-0.65ex,scale=0.4]
			\draw [black] (0,0) circle [];
			\filldraw[color=white!80, fill=white!15](0,0) circle (1);	
			\draw [black, thick]   (0,0) circle [radius=1.];
			\draw [black, thick, dashed]   (0,0) circle [radius=0.9];
			\begin{feynman}
				\vertex (A) at (0,2.);
				\vertex (C) at (0, 1.);
				\vertex (D) at (0, -1.);
%				\vertex (d) at (0.1,0) {\text{\footnotesize 1-loop }\normalsize} ;
				\vertex (B) at (0,-2.);
				\diagram*{
					(A) -- [photon] (C),
					(A) --[ charged scalar] (C),
					(D) -- [photon] (B),
					(D) --[ charged scalar] (B)
				};
			\end{feynman}
	\end{tikzpicture} } + 
\mathord{
\begin{tikzpicture}[baseline=-0.65ex,scale=0.4]
	\draw [black] (0,0) circle [radius=2cm];
	\begin{feynman}
		\vertex (A) at (-0.75,1.8);		
		\vertex (B) at (-0.75,-1.8);
		\vertex (C) at (0.75,1.8);
		\vertex (D) at (0.75,-1.8);
		\diagram*{
			(A) -- [photon] (B),
			(A) --[ fermion] (B),
			(C) -- [photon] (D),
			(C) --[ fermion] (D)
		};
	\end{feynman}
\end{tikzpicture} 
} -\epsilon\dfrac{\zeta(3)C_FN}{8\pi^2}+\ldots \ ,
\end{equation}
where in the first diagram we employed the double dashed/line of  (\ref{eq:difference theory}) to describe the one-loop propagators in the \textit{difference theory}. Finally, we find that 
\begin{equation}
	\label{eq:diagrammatic expression for Deltaw4}
	\begin{split}
		\mathord{
			\begin{tikzpicture}[radius=2.cm, baseline=-0.65ex,scale=0.4]
				\draw [black] (0,0) circle [];
				%\filldraw[color=gray!80, fill=gray!15](0,0) circle (1);	
				\draw [black, thick]   (0,0) circle [radius=1.];
				\draw [black, thick, dashed]   (0,0) circle [radius=0.9];
				\begin{feynman}
					\vertex (A) at (0,2.);
					\vertex (C) at (0, 1.);
					\vertex (D) at (0, -1.);
%					\vertex (d) at (0.1,0) {\text{\footnotesize 1-loop }\normalsize} ;
					\vertex (B) at (0,-2.);
					\diagram*{
						(A) -- [photon] (C),
						(A) --[ charged scalar] (C),
						(D) -- [photon] (B),
						(D) --[ charged scalar] (B)
					};
				\end{feynman}
		\end{tikzpicture} }= C_F \frac{\beta_0}{\epsilon(2\epsilon-1)} A_2(\epsilon)\ .
%	
%		 \dfrac{C_F}{8}
%	\left(\dfrac{\beta_0^\cR\Gamma^2(1-\epsilon)a_0(-2\epsilon)}{\pi^{-2\epsilon}\epsilon(2\epsilon-1)}\right) \ .		
	\end{split}
\end{equation}
Surprisingly, also the ladder-like diagrams provide an evanescent factor proportional to $\zeta(3)$. Using the well-known properties of the non-Abelian exponentiation of the Wilson loop operator, we obtain
\begin{equation}
	\label{eq:ladderg^4}
	\begin{split}
		\mathord{
		\begin{tikzpicture}[baseline=-0.65ex,scale=0.4]
			\draw [black] (0,0) circle [radius=2cm];
			\begin{feynman}
				\vertex (A) at (-0.75,1.8);		
						\vertex (B) at (-0.75,-1.8);
				\vertex (C) at (0.75,1.8);
				\vertex (D) at (0.75,-1.8);
				\diagram*{
					(A) -- [photon] (B),
					(A) --[ fermion] (B),
					(C) -- [photon] (D),
					(C) --[ fermion] (D)
				};
			\end{feynman}
		\end{tikzpicture} 
	} &=  \dfrac{C_F(2N^2-3)}{12N} A_1^2(\epsilon)
%	 \left(\dfrac{\Gamma(1-\epsilon)a_0(-\epsilon)}{8\pi^{-\epsilon}}\right)^2 \\ &
	- \epsilon \dfrac{C_F N\zeta(3)}{16\pi^2} \ .
\end{split}
\end{equation}
The evanescent $\zeta(3)$-like term results from a nested quadruple integration over the Wilson loop contour associated with the \textit{maximally non-Abelian} part of the diagram, namely the contribution of the diagrams characterized by the Casimir eigenvalues
% of the correction whose 
with the colour factor 
%only depends on 
$C_FC_{\rm adj} =C_F N$. Note that this combination again coincides with the colour factor associated with the single exchange diagrams (\ref{eq:ladder g2}).

\subsection{Three-loop corrections}
At order $\hat g_B^6$, we can use the fact that, up to evanescent corrections, all the interaction diagrams with internal line associated with vector-multiplet fields cancel the corrections resulting from  hypermultiplets in the adjoint representation. This is the statement that in the $\cN=4$ theory the observable only receives \textit{ladder-like} corrections, while in our case these cancellations reconstruct difference theory loop diagrams suggested by localization. We identify the following five classes of corrections:  
\begin{equation}
	\begin{split}
	\label{eq:diagrams at three-loop}
	W_6 & = \mathord{
	\begin{tikzpicture}[baseline=-0.65ex,scale=0.47]
		\draw [black] (0,0) circle [radius=2cm];
		\begin{feynman}
			\vertex (A) at (-0.75,1.8);		
			\vertex (B) at (-0.75,-1.8);
			\vertex (C) at (0.75,1.8);
			\vertex (D) at (0.75,-1.8);
			\vertex (O) at (0.,-2);
			\vertex (O1) at (0.,2);
			\diagram*{
				(A) -- [photon] (B),
				(A) --[ fermion] (B),
				(C) -- [photon] (D),
				(C) --[ fermion] (D),
				(O1)--[fermion] (O),
				(O1)--[photon] (O),
			};
		\end{feynman}
	\end{tikzpicture} 
} \ + \ \mathord{
		\begin{tikzpicture}[radius=2.cm, baseline=-0.65ex,scale=0.47]
			\draw [black] (0,0) circle [];
			\filldraw[color=gray!80, fill=gray!15](0,0) circle (1);	
			\draw [black, thick]   (0,0) circle [radius=1.];
			\draw [black, thick, dashed]   (0,0) circle [radius=0.9];
			\begin{feynman}
				\vertex (A) at (0,2.);
				\vertex (C) at (0, 1.);
				\vertex (D) at (0, -1.);
				\vertex (d) at (0.1,0) {\text{\footnotesize 2-loop }\normalsize} ;
				\vertex (B) at (0,-2.);
				\diagram*{
					(A) -- [photon] (C),
					(A) --[ charged scalar] (C),
					(D) -- [photon] (B),
					(D) --[ charged scalar] (B)
				};
			\end{feynman}
	\end{tikzpicture} } \ +	\mathord{
		\begin{tikzpicture}[radius=2.cm, scale=0.47, baseline=-0.65ex]
			\begin{feynman}
				\vertex (A) at (0,2.);
				\vertex (C) at (0,0);
				\vertex (D) at (-1.5, -1.3);
				\vertex (B) at (1.5, -1.3);
				\vertex (B1) at (1.,-0.8);
				\vertex (B2) at (0.5,-0.5);
				\diagram*{
					(A) -- [photon] (C),
					(A) --[ fermion] (C),
					(C) -- [photon] (B2),
					(C) -- [plain] (B2),
					(B) --[photon] (B1),
					(B) --[plain] (B1),
					(C) --[ fermion] (D),
					(C) --[ photon] (D)
				};
			\end{feynman}
			\draw [black] (0,0) circle [];
			\filldraw[color=white!80, fill=white!15](0.75,-0.65) circle (0.73);	
			\draw [black, thick, dashed] (0.75,-0.65) circle [radius=0.66cm];
			\draw [black] (0.75,-0.65) circle [radius=0.73cm];
%			\draw (-0.15,-0.7) node[anchor=west]{$\tiny\text{1-loop}$};
		\end{tikzpicture} 
	} \\[0.4em]
&  + 	\mathord{
		\begin{tikzpicture}[radius=2.cm, baseline=-0.65ex,scale=0.47]
			\draw [black] (0,0) circle [];
			\draw [black] (0,0) circle [radius=0.8cm];
			\draw [black, dashed,thick] (0,0) circle [radius=0.7cm];
			\begin{feynman}
				\vertex (A) at (0,2);
				\vertex (C) at (0,0.8);
				\vertex (D) at (-1.5, -1.3);
				\vertex (B) at (-0.7, -0.4);
				\vertex (B1) at (0.7,-0.4);
				\vertex (B2) at (1.5,-1.3);
				\diagram*{
					(A) -- [photon] (C),
					%						(A) --[ charged scalar] (C),
					(B) --[ anti fermion] (D),
					(B) --[photon] (D),
					(B1) --[fermion] (B2),
					(B1) --[photon] (B2),
				};
			\end{feynman}
%			\draw (-0.9,-0.) node[anchor=west]{$\tiny\text{1-loop}$};
		\end{tikzpicture} 
	} \ +	\mathord{
		\begin{tikzpicture}[baseline=-0.65ex,scale=0.47]
			\draw [black] (0,0) circle [radius=2cm];
			\draw [black] (-0.75,0) circle [radius=0.45cm];
			%\draw [black] (-0.75,0) circle [radius=0.5cm];
			%\filldraw[color=black!60, fill=black!5, very thick](-0.75,0) circle (0.5);
			\begin{feynman}
				\vertex (A) at (-0.75,1.8);
				\vertex (b) at (-0.75,0.45);
				\vertex (d) at (-0.75,-0.45);
				\vertex (B) at (-0.75,-1.8);
				\vertex (C) at (0.75,1.8);
				\vertex (D) at (0.75,-1.8);
				\diagram*{
					(A) -- [photon] (b),
					(A) --[ fermion] (b),
					(d) --[fermion] (B),
					(d) --[photon] (B),
					(C) -- [photon] (D),
					(C) --[fermion] (D)
				};
			\end{feynman}
		\filldraw[color=white!80, fill=white!15](-0.75,0) circle (0.73);	
			\draw [black, thick, dashed] (-0.75,0) circle [radius=0.66cm];
			\draw [black] (-0.75,0) circle [radius=0.73cm];
%			\draw (-1.65,-0.05) node[anchor=west]{$\tiny\text{1-loop}$};
		\end{tikzpicture} 
	}  \ +\cO(\epsilon) \ .
\end{split}
\end{equation}  
The  $\cO(\epsilon)$ terms are analogous to those we encountered in eq. (\ref{eq:difference}) and, in an analogous way, they could yield finite four-loop corrections whose analysis is beyond our current goal. As we will see in the following section, the diagrams in (\ref{eq:diagrams at three-loop})  guarantee the correct renormalization properties of the Wilson loop \cite{Dotsenko:1979wb,Brandt:1981kf,Korchemsky:1987wg}. The calculation of these contributions is extremely technical and will be examined in detail in an upcoming work \cite{Billo:2024}.  We find 
\begin{equation}
	\label{eq:result three-loop}
	\begin{split}
	W_6 &= C_F\dfrac{N^4-3N^2+3}{4608 N^2} + \dfrac{3\zeta(3)\cK_4^{\prime}}{2^8\pi^4N}
%	\\[0.3em]
%	& 
	+ C_F %\dfrac{(\beta_0)^2\Gamma(1-3\epsilon)\Gamma^2(\epsilon)\Gamma(1-\epsilon)}{\Gamma^2(2-2\epsilon)\Gamma(1+2\epsilon)}
	\dfrac{(\beta_0)^2}{(2\epsilon^2-\epsilon)^2}  
	A_3\\[0.3em]
	+ &  \dfrac{C_F (2N^2-3)}{6N} \dfrac{\beta_0 }{2\epsilon^2-\epsilon} A_1\, A_2  
%	\\[0.3em]	
%	& 
	+ \dfrac{7 C_F N\beta_0 \zeta(3)}{16\pi^2}  +  \cO(\epsilon)\ . 
	\end{split}
\end{equation}
Note that at this perturbative order we generated two terms proportional to $\zeta(3)$. The first one, which involves the quantity $\cK_4^\prime$, arises from the single-exchange diagrams dressed with the two-loop corrections to the adjoint scalar and gauge field propagators in the difference method, i.e. the first class of diagrams in (\ref{eq:diagrams at three-loop}). This term was originally studied  in \cite{Andree:2010na} and arises from a well-known Feynman integral which is regular when $\epsilon\to 0$ and proportional to $\zeta(3)$.  Being a finite and massless integral in four dimensions,  it retains the same form  and the same value on the sphere and in flat space.   

%Among these corrections, we  encounter the well-known  Feynman integral 
%\cite{Andree:2010na} 
%	\begin{equation}
%		\begin{split}
%			\mathord{
%				\begin{tikzpicture}[scale=0.7, baseline=-0.65ex]
%					\draw [black,thick] (0,0) circle [radius=1cm];
%					\draw [thick] (0,1) -- (0,-1);
%					\filldraw[black,thick] (1,0) circle (2pt);
%					\filldraw[black,thick] (-1,0) circle (2pt);
%					\filldraw[black,thick] (0,1) circle (2pt);
%					\filldraw[black,thick] (0,-1) circle (2pt);
%				\end{tikzpicture} 
%			} &= \int  \dfrac{\dd^4k}{(2\pi)^4}  \dfrac{\dd^4l}{(2\pi)^4}
%		\dfrac{  1 }{(k+p)^2(l+p)^2 k^2 l^2 (l-k)^2}\\
%		&=\dfrac{6\zeta(3)}{\left(4\pi\right)^4p^2} \ 
%	\end{split}
%\end{equation}  
%which, in configuration space, takes the same form in flat space and on the sphere. The topology of this diagram coincides with that of the second matrix model diagram in (\ref{eq:matrix model diagrams}) using the interpretation in (\ref{eq:difference theory}). 

The second contribution proportional to $\zeta(3)$ in eq. (\ref{eq:result three-loop}) is characterized by the same    colour factor predicted by the matrix model in eq. (\ref{eq:localization prediction}) and results from the last three classes of diagrams depicted in eq. (\ref{eq:diagrams at three-loop}). 

\section{Renormalization and comparison with the localization approach}
The dimensionally regularized Wilson loop v.e.v. $W$  
is ultraviolet divergent and must be renormalized in order to obtain a finite result. Since the operator is defined over a smooth contour, the divergences are removed just by the charge renormalization of the coupling $g_B$ \cite{Dotsenko:1979wb,Brandt:1981kf,Korchemsky:1987wg} which, in terms of $\hat g_B$,  amounts to  
\begin{equation}
	\label{eq:renormalized coupling}
	\hat g_B= (MR)^\epsilon\, g_*\,Z_{g_*}(\epsilon) \ .
\end{equation}
Here $g_*(M)\equiv g_*$ is the renormalized coupling at the renormalization scale $M$,  
%In the previous expression, $\mu$ is the renormalization scale, $g_*$ is the renormalized coupling, 
while $Z_{g_*}$ denotes the subtraction terms. The one-loop exactness of the $\beta$-function (\ref{beta0}) implies that in the MS scheme we find the following expression for the subtraction terms 
\begin{equation}
	\label{eq:subtraction terms}
Z_{g_*}^2(\epsilon)=
%\dfrac{1}{1-\dfrac{\beta_0 g_*^2}{\epsilon}} \ .
\left(1-\dfrac{\beta_0 g_*^2}{\epsilon}\right)^{-1}\ .
\end{equation}
Inserting eq. (\ref{eq:renormalized coupling}) in the explicit expression (\ref{eq:perturbative}) of $W$ that follows from the previous results for $W_{2,4,6}$ all the (UV) divergences cancel and we can define the renormalized observable as
\begin{equation}
W_*=\lim_{\epsilon\to 0} W \ .
\end{equation}  

When $\epsilon\to 0$, the overall dependence  on the scale $M$ disappears and $W_*$ satisfies
the usual Callan-Symanzik equation \cite{Billo:2023igr}.
%\begin{equation}
%\left(M\dfrac{\partial}{\partial M}+\beta(g_*)\dfrac{\partial}{\partial g_*}\right)W_*
%%_{\mathsf{R}}\left(RM,g_*\right) 
%=0 \ .
%\end{equation} 
This implies that $W_*$ must actually depend on $M$, $g_*$ and $R$ through the running coupling $g$ defined in (\ref{eq:running coupling}). This is in fact what happens; moreover, the explicit expression of $W_*(g)$ is quite simple and 
%in fact it 
coincides perfectly with the localization result in (\ref{eq:localization prediction}):
\begin{equation}
	\label{WiscW}
		W_*(g) = \cW(g) + \cO(g^8)
\end{equation}
in the regime of validity
% of the matrix model result, see 
specified in (\ref{eq:range}).  In fact, only when $R M \gg 1$ the $\log RM$-terms, which  arise when we replace the bare coupling with renormalized one (\ref{eq:renormalized coupling}),  dominate over other scheme-dependent terms which we can then neglect to obtain the relation (\ref{WiscW}). 
Potentially, the presence of these large logarithmic contributions could make perturbation theory ill-defined. However,  when $R \Lambda \ll 1$ we can resum these large logarithms in the effective coupling $g$ which remains small. Beyond the range (\ref{eq:range}), we expect that the 
%perturbative expansion 
results on $\mathbb{S}^4$ differ from those in flat space by 
%infrared 
power-like corrections proportional to $R\Lambda$ -- see the discussion after eq. (\ref{eq:range}).

Some further comments are in order. Firstly, we remark the crucial role of the evanescent factors in the two-loop corrections (\ref{eq:difference}) and (\ref{eq:ladderg^4}). Upon renormalization, these factors interfere with the UV poles of the bare coupling $\hat{g}_B$ (\ref{eq:subtraction terms}) and provide finite (three-loop) corrections proportional to $\beta_0\zeta(3)$ which combine with the analogous ones in (\ref{eq:result three-loop}).   However, 
it turns out that the $\zeta(3)$-terms resulting from the \textit{Mercedes/lifesaver-like} diagrams, namely the correction depicted in (\ref{eq:difference}) and the analogous ones in (\ref{eq:diagrams at three-loop}), do not contribute to the final result. Thus, in the perturbative field approach only the $\zeta(3)$-corrections resulting from the (two/three-loop) double-exchange diagrams are relevant an reproduce the matrix model prediction. Importantly, this effect ties nicely in with the diagrammatic approach of the matrix model (\ref{eq:matrix model diagrams}). The reason is that the $\zeta(3)$-like part of the multiple-exchange corrections emerges from the \textit{maximally non-Abelian} parts of the diagrams which, being proportional to $C_F$, behave as single-exchange diagrams in agreement with the  matrix model prediction (\ref{eq:matrix model diagrams}). 

Let us also note that, once re-expressed in terms of the running coupling, the renormalized v.e.v. up to three-loop order can be described in terms of few diagrams. Beside the ladder corrections, there is the irreducible part of the three-loop single exchange -- the second diagram in (\ref{eq:diagrams at three-loop}) -- which is
characterized by the colour factor $\cK^\prime_4$ and is  present also in the superconformal cases
\cite{Andree:2010na, Billo:2019fbi} and the term proportional to $\beta_0 \zeta(3)$ that arises from a pinching limit of the maximally non-Abelian part of the double exchange ladder diagram.  
%The correspondence of this last term with the matrix model diagrams depicted in (\ref{eq:matrix model diagrams}) is not obvious, but it must be kept in mind that in the matrix model the vertices are not positioned in space-time 

\section{Conclusions and future perspectives}
%In this letter, w
We examined the relation between supersymmetric localization and standard perturbative techniques in flat space for generic $\mathcal{N}=2$ SYM theories with non-vanishing $\beta$-function.

We studied via localization to a matrix model the vacuum expectation value of the one-half BPS Wilson loop on $\mathbb{S}^4$. Within the regime described in (\ref{eq:range}), we showed that the matrix model predictions match standard perturbation theory based on Feynman diagrams techniques in flat space up to order $g^6$.  We  precisely related the matrix-model effective diagrams associated with the $\zeta(3)$ terms to the flat-space preturbative expansion. Our results not only provide a
%an highly 
non-trivial test of the localization approach for non-conformal theories but also unveil the subtle reorganization of the conventional Feynman diagrams into the matrix-model average. It would be interesting to extend our analysis to the next perturbative order and try to generalize the understanding at all loops. Another natural investigation 
%to perform, along these lines, 
would 
%consist in examining 
be to examine correlators of local operators: 
%supersymmetric 
localization gives exact results, in the non-conformal case, also for classes of two-point functions that can be compared with flat-space perturbation theory \cite{Billo:2019job} . It would be interesting to reanalyse these observables at the light of the present computations. Exact all-orders expressions on $\mathbb{S}^4$ have been also used to study the large-order behaviour of the perturbative series, in connection with resurgent techniques \cite{Aniceto:2014hoa}, for different $\mathcal{N}=2$ SYM theories. Reconsider the non-conformal case and its relation with a flat-space analysis could further improve our understanding of the perturbative results and their gauge-invariant resummation. 
\begin{acknowledgments}
\emph{Acknowledgments.} We thank Grisha Korchemsky for carefully reading the  manuscript and for many illuminating discussions, Francesco Galvagno, Alberto Lerda, Marialuisa Frau and Igor Pesando for lively exchange of ideas. A.T. is grateful to the Institut de Physique Théorique (CEA) for the kind hospitality when essential part of this work was done.
\end{acknowledgments}

%\nocite{*}
%\bibliography{Biblio_Def}
\providecommand{\noopsort}[1]{}\providecommand{\singleletter}[1]{#1}%

\end{document}